# Large unidirectional spin Hall and Rashba-Edelstein magnetoresistance in topological insulator/magnetic insulator heterostructures

Yang Lv[1*], James Kally[2*], Tao Liu[3*], Protyush Sahu[4], Mingzhong Wu[3], Nitin Samarth[2] and Jian-Ping Wang[14#]

[1]Department of Electrical and Computer Engineering, University of Minnesota, Minneapolis, Minnesota 55455, USA

[2]Department of Physics, The Pennsylvania State University, University Park, Pennsylvania 16802, USA

[3]Department of Physics, Colorado State University, Fort Collins, Colorado 80523, USA

[4]School of Physics and Astronomy, University of Minnesota, Minneapolis, Minnesota 55455, USA

[*]equal contribution

[#]E-mails: jpwang@umn.edu;

**Thanks to its unique symmetry, the unidirectional spin Hall and Rashba-Edelstein magnetoresistance (USRMR) is of great fundamental and practical interest, particularly in the context of reading magnetization states in two-terminal spin-orbit torque switching memory and logic devices. Recent studies show that topological insulators could improve USRMR amplitude. However, the topological insulator device configurations studied so far in this context, namely ferromagnetic metal/topological insulator bilayers and magnetically doped topological insulators, suffer from current shunting by the metallic layer and low Curie temperature, respectively. Here, we report large USRMR in a new material category - magnetic insulator/topological insulator bi-layered heterostructures. Such structures exhibit USRMR that is about an order of magnitude larger than the highest values reported so far in all-metal Ta/Co bilayers. We also demonstrate current-induced magnetization switching aided by an Oersted field, and electrical read out by the USRMR, as a prototype memory device.**

The recently discovered unidirectional spin Hall magnetoresistance (USMR)[1–3] is very interesting from both scientific and engineering perspectives. In a bilayer structure consisting of a ferromagnetic (FM) layer and a non-magnetic (NM) heavy metal layer, the USMR originates from interactions between the spins at the interface generated by the spin Hall effect (SHE) in the NM and spin conduction channels in the FM; its strength is proportional to the projection of the magnetization in the FM along the direction of spin polarization at the FM/NM interface. From an engineering perspective, the USMR is particularly interesting since it possesses a unique symmetry that is sensitive to 180-degree magnetization changes. In the spin-orbit torque (SOT) switching geometry, an FM layer can be switched by applying a current through the spin Hall channel[4]. However, for reading the state of the FM, a full magnetic tunneling junction (MTJ) structure has to be built, and an additional terminal needs to be added to the device[4]. Using the USMR effect, such switching devices no longer require an additional MTJ structure and an extra terminal for reading.

Despite these advantages, the USMR demonstrated so far is still too small to be practical. In a FM/NM system, the amplitude of the USMR peaks at the NM thickness of about a few times of spin diffusion length[1,5] and is ultimately limited by the spin Hall angle of the NM, which characterizes the capability of converting a charge current into a transverse spin current. Topological insulators (TIs) are a class of materials whose bulk is ideally electrically insulating while the surfaces are conductive[6–9]. Electrons on the surfaces of a TI are naturally spin-polarized due to "spin-momentum locking" (SML). In order to reflect the fact that the spin generation relies partially on the topological surface states, the term unidirectional spin Hall and Rashba−Edelstein magnetoresistance (USRMR) is used in systems involving TIs[10]. Thanks to the TI's high efficiency in converting surface and bulk charge currents into spins even at room temperature[11,12], it is natural to expect that the USRMR in FM/TI systems may be much larger than the USMR in FM/NM structures. Indeed, recent measurements of $Bi_2Se_3$/CoFeB demonstrate this expectation[10]. However, the metallic FM layer in such structures shunts most of the current injected, and thereby makes the current flowing in the TI low and fails to generate significant spin accumulation at the interface. In this work, we take one step further to investigate the USRMR behavior of a magnetic insulator (MI)/TI system that consists of an yttrium iron garnet ($Y_3Fe_5O_{12}$, YIG)/$Bi_2Se_3$ (BS) bi-layered structure. Our work presents the first observation of the room-temperature USRMR in the MI/TI system. At 150 K, we observe a record amplitude of the USRMR per current density per total resistance, about 5 times larger than that in the metallic CoFeB/$Bi_2Se_3$ system[10] and an order of magnitude larger than the highest values reported in all-metal Ta/Co bilayers[1]. Notably, this is a large effect even in non-ideal TI films ($Bi_2Se_3$) with both bulk and surface conduction.

On the other hand, extensive studies have explored the paths to achieve current-induced magnetization switching with TIs in a variety of materials combinations: TI/FM bilayers[13,14], TI/ferrimagnet bilayers[15], magnetic TI/TI bilayers[16] and NM/perpendicular MI bilayers[17]. More excitingly, in this paper, we also demonstrate current-induced magnetization switching in TI/MI bilayers, aided by an Oersted field, and electrical read out by the USRMR, as a prototype memory device. This also could be a first step to implement spin switch concept that was proposed based on topological insulator material[18].

The phenomenon of the USRMR in a YIG/BS bilayer is sketched in Fig. 1. In the presence of a charge current, *j*, in the BS layer, spins are generated at the YIG/BS interface due to the SML as well as spin-orbit coupling (SOC) in the bulk. The magnetic proximity effect of the YIG induces ferromagnetism at the YIG/TI interface[19] and alters the scattering of the spins generated by the TI. Depending on the relative directions between the spin polarization of electrons at the interface and the magnetization in the YIG, different resistance is induced because of spin-dependent scattering. We observed the USRMR effect at temperatures between 70 K and 300 K in YIG(30 nm)/BS($t$)/Al(4 nm) structures with $t$=6, 8, and 15 QL. The YIG layer is deposited on a single-crystal (111) gadolinium gallium garnet ($Gd_3Ga_5O_{12}$, GGG) substrate in an ultrahigh vacuum sputtering system at room temperature first and is then annealed in-situ in $O_2$ at 800° C. Following the YIG deposition, the BS layer and the Al capping layer are grown in turn by molecular beam epitaxy (MBE). Hall bars with nominal length $L$=50 μm and width $W$=10 or 20 μm are fabricated by standard photolithography processes and tested with harmonic measurements in both longitudinal and transverse resistance configurations. Note that we will refer all samples as YIG/BS($t$=6, 8 or 15) from now on for simplicity.

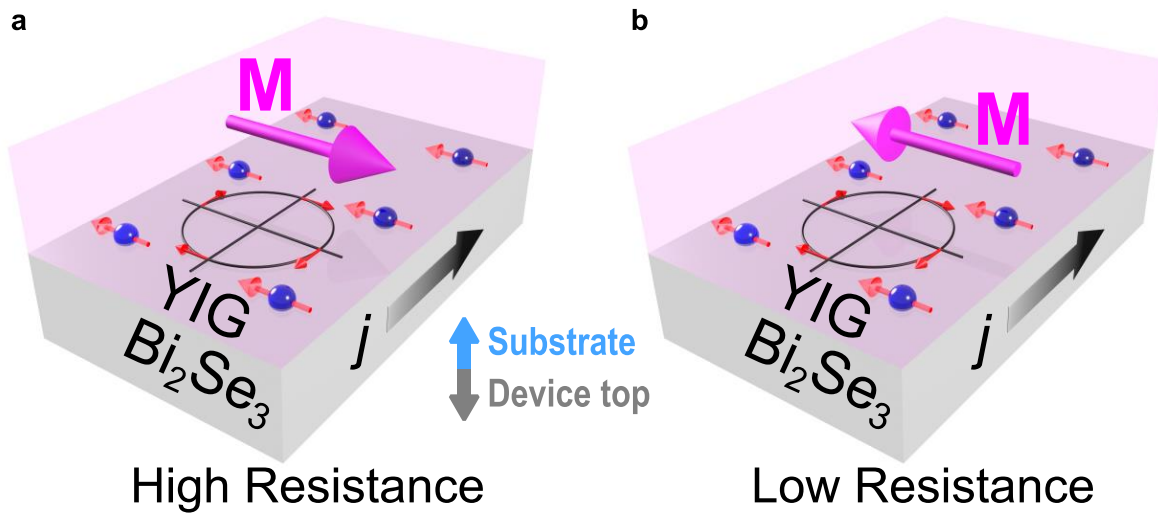

**Fig. 1** Illustration of USRMR in an MI/TI bilayer. Spins are generated at interface when a charge current is applied. The relative direction of spins to magnetization of either **a** parallel or **b** anti-parallel results in different resistance state.

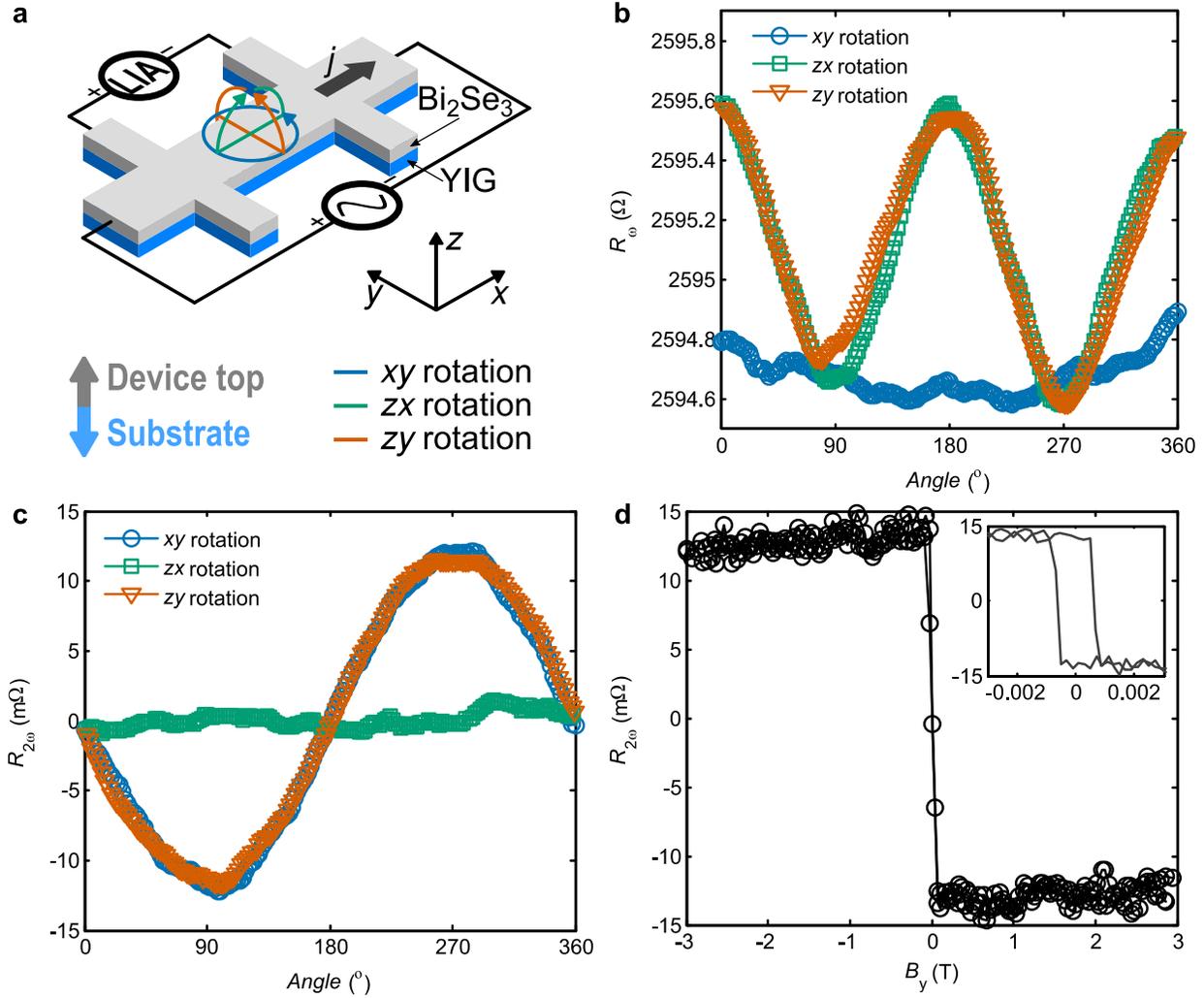

**Fig. 2** Angular and field dependence of longitudinal resistance. **a** Longitudinal resistance measurement setup and definitions of rotation planes. First harmonic **b** and second harmonic **c** resistances of the YIG(30 nm)/BS(8 QL)/Al(4 nm) sample at 150 K with $L$=50 μm and $W$=20 μm are shown when the external field is rotated in three orthogonal planes. The starting points and zero angles are at $+x$, $+z$ and $+z$, the directions of rotation of increasing angle are $+x$ to $+y$, $+z$ to $+x$ and $+z$ to $+y$, for $xy$, $zx$ and $zy$ rotations, respectively. d) Second harmonic resistance vs. external field swept along $y$-axis of the same sample at 150 K.

Figure 2a shows the longitudinal resistance measurement setup and the definitions of the coordinates and rotation planes. Note that the sample in this figure is 'up-side-down' compared to the sketches in Fig. 1. The "+" and "-" signs and the arrow of $j$ indicate the relative polarities of current source outputs and lock-in amplifier (LIA) inputs. Zero angles are at the +*x*, +*z*, and +*z* directions for the *xy*, *zx*, and *zy* rotations, respectively. The rotation directions for increasing angles are indicated by the arrows in the center area of the Hall bar. An external field of 3 T is applied and is rotated in the *xy*, *zx*, and *zy* device planes while the first order resistance $R_\omega$ and the second harmonic resistance $R_{2\omega}$ are recorded using a sinusoidal A.C. current with a frequency of $\omega/2\pi=33$ Hz and a peak amplitude of 707.1 µA. Figures 2b and 2c show the angle dependencies of $R_\omega$ and $R_{2\omega}$, respectively, of the YIG/BS(8) sample measured at 150 K for *L*=50 µm and *W*=20 µm. One can see that $R_\omega$ exhibits a behavior that differs from both the conventional anisotropic magnetoresistance and the spin Hall magnetoresistance since it varies only when the angle between the +*z* direction and the applied field changes. This is the result of the parabolic magnetoresistance of the TI. In stark contrast, $R_{2\omega}$ varies with the angle between the +*y* direction and the magnetization, similar to the behavior observed previously in NM/FM bilayers[1]. One can clearly see that in Fig. 2c $R_{2\omega}$ varies with a period of exactly 360° for the *xy* and *zy* rotations but shows a flat response for the *zx* rotation. The amplitude of $R_{2\omega}$ is about 13.66±0.14 mΩ at an average current density of 0.4419 MA cm$^{-2}$. Fig. 2d shows the $R_{2\omega}$ vs. magnetic field applied along *y* direction, $B_y$. The $R_{2\omega}$ does not change significantly further up to 3 T once the magnetization of YIG is switched around a few mT. This is an indication that the $R_{2\omega}$ signal observed is unlikely the unidirectional magnetoresistance (UMR) which was recently discovered in magnetic TI/TI bilayer systems, since UMR decreases significantly at high field due to a decrease of the magnon population[20]. Note that during the measurements, $R_{2\omega}$ is read from the Y-channel of the LIA at a frequency of 2$\omega$. Due to the nature of the harmonic measurement scheme and LIA, the reading is in fact -1/2 of the actual resistance change that would have been observed directly with a DC current. In other words, $R_{2\omega}<0$ and $R_{2\omega}>0$ represent higher and lower resistances, respectively, under a forward current bias ($j>0$).

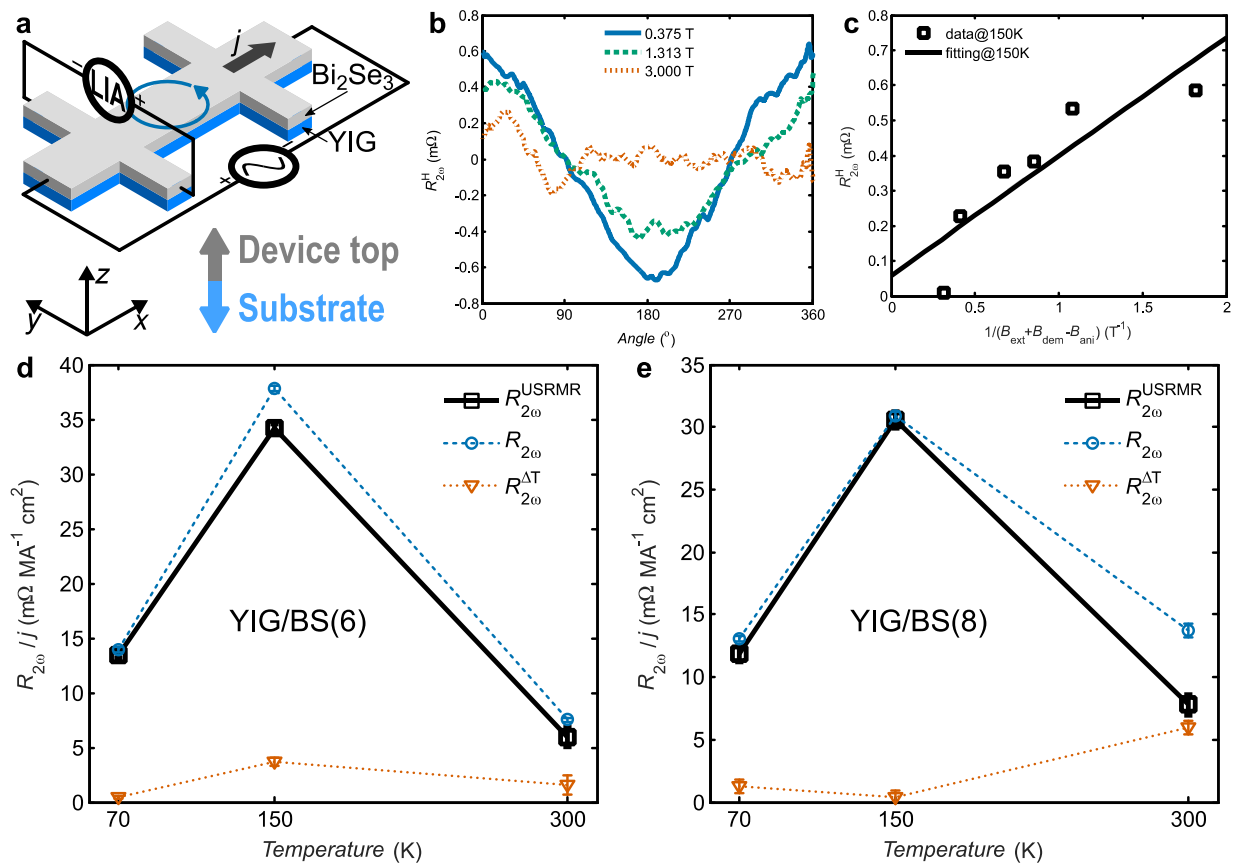

**Fig. 3** Second harmonic Hall resistance measurement setup and results. **a** Transverse/Hall resistance measurement setup. **b** Examples of second harmonic Hall resistance of YIG(30 nm)/BS(8 QL)/Al(4 nm) sample at 150 K with $L$=50 μm and $W$=20 μm vs. angle in $xy$ plane rotation with 20 mT and 3 T external fields. **c** Second harmonic Hall resistance measured with various external fields are plotted vs. reciprocal of total field and linear fitted. The intercept of the fitted line represents the contribution of ANE/SSE. The longitudinal $R_{2\omega}$ consists of the contribution of the thermoelectric effects $R_{2\omega}^{\Delta T}$ and USRMR $R_{2\omega}^{USRMR}$. Each component vs. temperature is plotted for **d** YIG(30 nm)/BS(6 QL)/Al(4 nm) sample and **e** YIG(30 nm)/BS(8 QL)/Al(4 nm) sample. Data of the two samples under 300 K is obtained with total current density of 1.0607 MA cm$^{-2}$ and a resistor attenuation circuit in front of LIAs to improve signal-to-noise ratio. The rest of the data is obtained with 0.4419 MA cm$^{-2}$. The whole plots are normalized by current density. The error bar in **d** and **e** indicates uncertainty bounds with 95% confidence. The uncertainties reflect the variations of observed signal level in the field sweep and angle rotation.

Figure 3a shows the Hall resistance measurement setup. The transverse Hall resistance is measured while the external field is rotated in the $xy$ plane. Figure 3b presents the second-harmonic Hall resistance $R_{2\omega}^{H}$ vs. angle responses measured with the external fields of different strengths, as indicated. The amplitude of $R_{2\omega}^{H}$ is much smaller than that of $R_{2\omega}$. If UMR were dominant, it would have yielded $R_{2\omega}^{H}$ signals that is 1/3 of the amplitude of the $R_{2\omega}$ under the same external field strength in the $xy$ rotation[16]. Therefore, based on this observation and that shown in fig. 2d, we adopt the picture of the USMR effect to analyze our data.

Due to Joule heating in the device while passing current for testing, both the anomalous Nernst effect (ANE) and the spin Seebeck effect (SSE) contribute to the $R_{2\omega}$ signal. To separate their contributions (denoted as $R_{2\omega}^{\Delta T}$) from the USRMR contribution, we carried out a series of measurements of $R_{2\omega}^{H}$ for the $xy$-plane rotation in various external field strengths. The $R_{2\omega}^{H}$ contains contributions from not only the ANE and the SSE, but also the magnetization oscillations of YIG induced by field-like (FL) and damping-like (DL) SOTs. The contributions from the ANE, the SSE, and the DL SOT are all proportional to $cos\varphi$ while that from the FL SOT is proportional to $cos3\varphi + cos\varphi$ (ref [21]), where $\varphi$ is the angle of magnetization from $+x$ direction in an $xy$-plane rotation and follows the direction of increasing angle indicated in Fig. 3a. Since the DL SOT and the FL SOT contribute to $R_{2\omega}^{H}$ by perturbing the magnetization in the YIG, their effects diminish when the external field is very large. Figure 3b shows that the data measured with external field $B_{ext}$=375 mT contains a dominant $cos\varphi$ contribution and a small $cos3\varphi$ component, while with $B_{ext}$=3 T the data exhibits almost no angle dependence. We first obtain $R_{2\omega}^{H,\Delta T}$ by fitting the angle dependent data, allowing us to extract the amplitudes of the $cos\varphi$ and $cos3\varphi$ components. The FL SOT contribution can then be easily determined and separated. This leaves the contributions of ANE/SSE and DL SOT. We plot the data corresponding to these contributions versus the reciprocal of total field, as shown in Fig. 3c. In this figure, $B_{demag}$-$B_{ani}$ is the demagnetization field minus the perpendicular anisotropic field of the MI layer, which is about 176 mT[22]. Since the effect of the DL SOT will diminish at infinite field, the intercept of the fitted line is the contribution of ANE/SSE to the 2$^{nd}$ order Hall resistance. Then, the contribution of ANE/SSE to the longitudinal resistance $R_{2\omega}$, $R_{2\omega}^{\Delta T}$, is obtained by scaling that from the Hall resistance, $R_{2\omega}^{H,\Delta T}$, with a factor of device's aspect ratio. Finally, the USRMR is determined once the ANE/SSE contribution is subtracted from the total $R_{2\omega}$ signal.

Figures 3d and 3e show the $R_{2\omega}$, $R_{2\omega}^{\Delta T}$ and $R_{2\omega}^{USRMR}$ of the YIG/BS(6) sample and YIG/BS(8) sample, respectively, at various temperatures. The current densities used are 1.0607 MA cm$^{-2}$ and 0.4419 MA cm$^{-2}$ at 300 K and other lower temperatures, respectively. The dimensions of the Hall bars for all data points in Fig. 3d and 3e are $L$=50 μm and $W$=20 μm, except for the one of the Hall bar tested of YIG/BS(8) at 300 K being $L$=50 μm and $W$=10 μm. The error bars in Fig. 3d and 3e indicate uncertainty bounds with 95% confidence. It is evident from the figures that the $R_{2\omega}^{\Delta T}$ is only a small portion of the total measured $R_{2\omega}$, except for at 300 K. Since temperature affects the chemical potential, the carrier concentration, and the relative contributions to transport from surface and bulk conduction in the TI, the charge-spin conversion of the TI is strongly temperature dependent. On the other hand, the magnetic behavior of the YIG, especially the effective damping, is also strongly temperature dependent. As a result, the USRMR is heavily temperature dependent. Both YIG/BS(6) and YIG/BS(8) samples show largest USRMR at 150 K and much less USRMR at 70 K and 300 K. This 'low-high-low' temperature profile of USRMR could originate from the competition between TI's charge-spin conversion and YIG's spin relaxation; the TI generates more spins at decreased temperature[23,24] while the YIG relaxes spin accumulation more severely at low temperatures[25,26]. Another observation is that at lower temperatures, both YIG/BS(6) and YIG/BS(8) samples show very little thermoelectric effects and the USRMR contributes to the most of the total $R_{2\omega}$ signal. While at 300 K, the thermoelectric effect is still weak in the YIG/BS(6) sample but becomes much more prominent in the YIG/BS(8) sample. This is possibly due to the 8 QL BS layer having larger resistivity and generating more heat than the 6 QL BS layer at high temperatures.

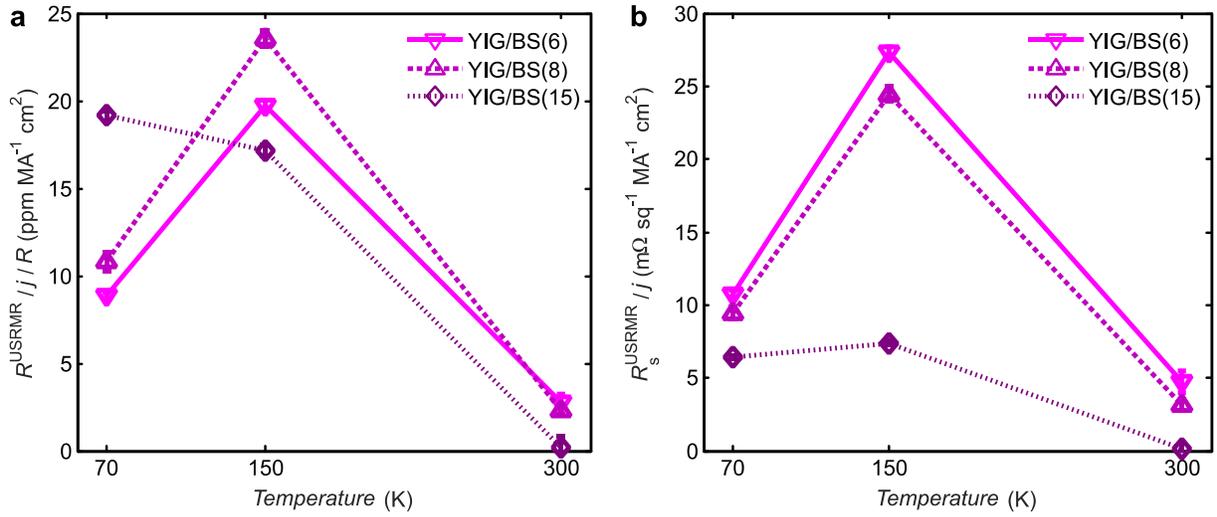

**Fig. 4** USRMR vs temperature for all samples. **a** USRMR per current density per total resistance and **b** sheet USRMR per current density of all four samples at various temperatures. The error bar indicates uncertainty bounds with 95% confidence. The uncertainties reflect the variations of observed signal level in the field sweep and angle rotation data.

Figure 4 shows USRMR per current density per total resistance ($R^{\text{USRMR}}/R/j$) and sheet USRMR per current density ($R_s^{\text{USRMR}}/j$) of all three samples as a function of temperature. The error bars indicate uncertainty bounds with 95% confidence. Note that $R^{\text{USRMR}}$ is defined as the amplitude of USRMR: $R^{\text{USRMR}} = \frac{1}{2}[R(\pm M, \pm j) - R(\pm M, \mp j)]$. And $R_s^{\text{USRMR}}$ is $R^{\text{USRMR}}$ normalized by device aspect ratio. We also would like to point out that the USRMR obtained by harmonic measurements, $R_{2\omega}^{\text{USRMR}}$, is half of the $R^{\text{USRMR}}$, by the nature of LIA and harmonic measurements. And the sheet USRMR, $R_s^{USRMR}$, is also doubled from the harmonic measurement results, $R_{s,2\omega}^{USRMR}$. The $R^{USRMR}/R/j$ and $R_s^{USRMR}/j$ provide more meaningful figure-of-merit for comparisons of USRMR across different thin film systems regardless of their testing currents and lateral dimensions. As shown in Fig. 4, these two values show very similar temperature dependent trends for the YIG/BS(6) and YIG/BS(8) samples. The largest $R^{USRMR}/R/j$ is found in the YIG/BS(8) sample at 150 K of 23.56±0.48 ppm MA$^{-1}$ cm$^2$ and it is about an order of magnitude larger than the largest USRMR reported in Ta/Co system (~1 ppm MA$^{-1}$ cm$^2$)[1]. And the largest $R_s^{USRMR}/j$ is found in YIG/BS(6) sample at 150 K of 27.31±0.38 mΩ sq$^{-1}$ MA$^{-1}$ cm$^2$. The difference of $R_s^{USRMR}/j$ between the YIG/BS(6) and YIG/BS(8) samples is almost constant across all temperatures despite the substantial change in their absolute values.

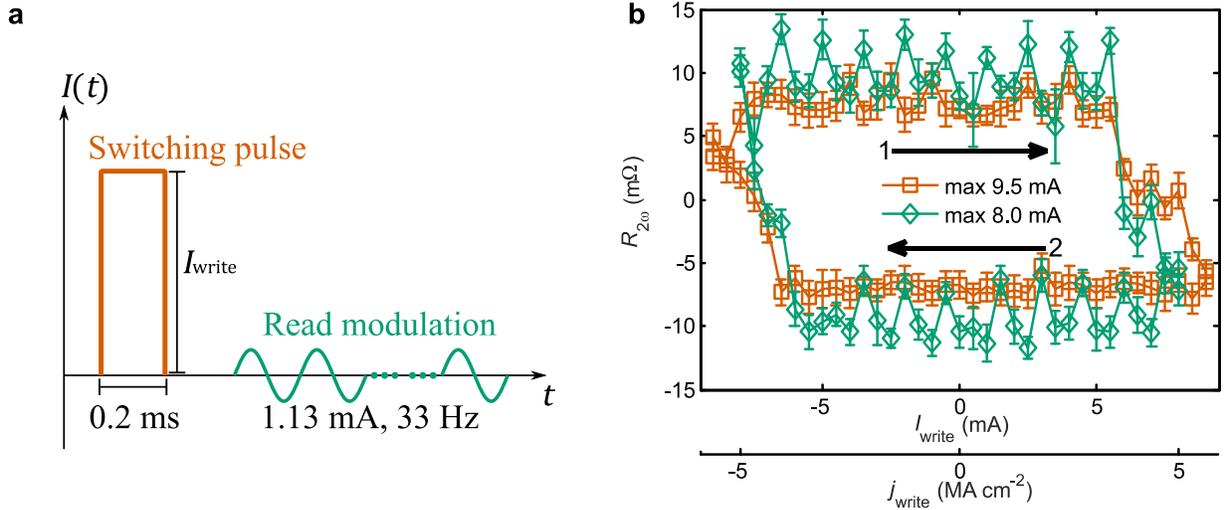

**Fig. 5** Demonstration of current-induced magnetization switching and read out by USRMR. **a** Illustration of testing sequence for pulse switching and USRMR reading measurements. The acquisition of each data point consists of a rectangular switching current pulse of 0.2 ms width and certain amplitude $I_{write}$ followed by a mild continuous sinusoidal A.C. current of 1.1312 mA amplitude and 33Hz, which allows the LIAs for reading $R_{2\omega}$. **b** The longitudinal second harmonic resistance $R_{2\omega}$ of YIG/BS(8) Hall bar sample vs. switching current pulse amplitude $I_{write}$ sweeps of maximum 8.0 mA (bluish green diamonds) and maximum 9.5 mA (orange squares) at 150 K. The secondary horizontal axis shows the equivalent current density of $I_{write}$.

The error bar indicates the range of $\pm\sigma$, where $\sigma$ is the standard deviation of 50 resistance readings from LIAs during a 10 s reading period.

Recent studies have demonstrated current-induced magnetization switching with TIs in a variety of materials combinations: TI/FM bilayers[13,14], TI/ferrimagnet bilayers[15], magnetic TI/TI bilayers[16] and NM/perpendicular MI bilayers[17]. However, current-induced magnetization switching of a MI using a TI has not yet been reported. Here, with the convenient aid of the USRMR, we demonstrate such in-plane magnetization switching with Hall bar devices. Figure 5a shows the testing sequence of the current switching and USRMR reading experiment. A switching pulse with an amplitude of $I_{write}$ and a width of 0.2 ms is applied to the device. Then, a moderate A.C current $I_{read}$ (1.1312 mA amplitude) is applied to allow the LIAs for reading the USRMR of the device, which tells the magnetization state. This cycle is repeated multiple times to complete a current switching hysteresis sweep. Figure 5b shows the $R_{2\omega}$ vs. $I_{write}$ of the YIG/BS(8) Hall bar sample of $L$=30 μm and $W$=20 μm at 150K. At large negative current, the USRMR ($R_{2\omega}$) is high, while at large positive current the USRMR ($R_{2\omega}$) is low. At $I_{write}$=0, the USRMR ($R_{2\omega}$) maintains two separate stable levels. This indicates that the magnetization of the YIG layer is manipulated by the charge current. The figure also indicates a critical switching current density smaller than 5.0 MA cm$^{-2}$ at 150 K, which is lower than the current density required to switch perpendicular MI with Pt at room temperature.[17] It is comparable to that needed to switch a FM with a TI at room temperature[14] and a perpendicular ferrimagnet with a TI at room temperature[15]. However, the coercivity of the device along the y-direction is much lower, being about 0.75 mT originally and at worst 0.35 mT after degradation. The Oersted field with 5.0 MA cm$^{-2}$ current density is about 0.25 mT. Therefore, although the experiment shows current-induced magnetization switching, the Oersted field contributes a sizeable portion of the switching.

In summary, we observed a nonlinear magnetoresistance that is sensitive to the magnetization component projected in the in-plane direction transverse to current direction in MI/TI (YIG/$Bi_2Se_3$) heterostructures. The harmonic Hall resistance measurements indicated that this magnetoresistance should be related with USMR or USRMR, instead of UMR. The USRMR in the MI/TI system was observable with a much lower current density compared to all metallic NM/FM bilayers. The large USRMR was attributed to the absence of current shunting by the insulating magnetic layer and better charge current utilization in the TI channel. Further, the temperature stability of the YIG makes the USRMR even present at room temperature, which is crucial for future applications. The largest USRMR is observed at 150 K and is about an order of magnitude larger than the best USMR reported in Ta/Co systems. Current-induced in-plane magnetization switching of YIG in Hall bar devices was demonstrated with the aid of USRMR. The large USRMR we observed in YIG/BS system should shed light on a new category of material system that is potentially an attractive platform for practical two-terminal SOT switching devices that can be read out by USRMR.

**Methods**

The YIG films were grown on GGG(111) substrates by RF magnetron sputtering at room temperature and then in-situ annealed at 800 °C for 2 hours under the oxygen pressure of 1 Torr, with a heating rate of 10 °C/min and a cooling rate of 2 °C/min.

After exposure to air, the YIG films were then transferred to an EPI 620 MBE for the Bi-chalcogenide deposition. $Bi_2Se_3$ films were grown from high purity (5N) Bi and Se evaporated from Knudsen cells at a beam equivalent pressure flux ratio of 1:14. The substrate temperature according to a radiatively coupled thermocouple was 325°C (pyrometer reading of 250°C) and the growth rate was 0.17 nm/min. The films have a root mean squared (RMS) roughness of approximately 0.7 nm over a 25 $\mu m^2$ area measured by atomic force microscopy (AFM). Film thickness was measured by X-ray reflectivity and crystal quality by high-resolution X-ray diffraction rocking curves of the (006) crystal plane. The latter show a full width half max (FWHM) of approximately 0.28 degrees. A 4-nm-thick capping layer of Al was grown before each sample left the MBE chamber.

The thin film stacks were then fabricated with a standard photolithography process followed by an ion milling etching step to define the Hall bars. Then, a second photolithography process and an e-beam evaporation followed by a liftoff step were performed to make electrode contacts.

The devices were tested in a Quantum Design PPMS which provides temperature control, external field and rotation. A sinusoidal A.C. current of 33 Hz was supplied by a Keithley 6221 current source. A Stanford Research SR830 or an EG&G 7265 LIA together with an EG&G 7260 LIA were used to measure the first and second harmonic voltages, respectively and simultaneously. For USRMR vs. switching current measurements, each acquisition of data point consists of a switching stage, in which the LIAs are broken from the device by a Keithley 7001 switch box and a rectangular current pulse is fired, and a reading stage, in which the LIAs are closed back to the device while a mild sinusoidal A.C. current is applied, and device voltage is read back. In the reading stage, the LIAs continuously recorded readings every 0.2 s during the period of between the 10th second and 20th second since the application of the sinusoidal A.C. current. Then the mean and standard deviation of reading samples are used for plots.

**Acknowledgements**

This work was supported in part by C-SPIN, one of six centers of STARnet, a Semiconductor Research Corporation program, sponsored by MARCO and DARPA. Parts of this work were carried out in the University of Minnesota Nanofabrication Center which receives partial support from NSF through NNCI program. We would also like to thank Timothy Peterson for his help on the usage of PPMS. The work at CSU was also supported by SHINES, an Energy Frontier Research Center funded by DOE, and NSF (EFMA1641989 and DMR-1407962).

**Author Contributions**

Y.L. and J.-P.W. conceived the research and designed the experiments. T.L. and M.W. grew and provided YIG thin films. J.K. and N.S. grew and provided BS thin films. Y.L. fabricated and measured devices and analyzed data. Y.L. developed and carried out the current switching measurements. P.S. helped with the current switching measurements. All authors reviewed and discussed results. Y.L. and J.-P.W. wrote the manuscript. All authors contributed to the completion of the manuscript.

ferromagnet/normal-metal bilayers. *Phys. Rev. B* **90,** 224427 (2014).

22. Wang, H. *et al.* Surface-state-dominated spin-charge current conversion in topological-tnsulator-ferromagnetic-insulator heterostructures. *Phys. Rev. Lett.* **117,** 76601 (2016).

23. Deorani, P. *et al.* Observation of inverse spin hall effect in bismuth selenide. *Phys. Rev. B* **90,** 94403 (2014).

24. Wang, Y. *et al.* Topological surface states originated spin-orbit torques in Bi2Se3. *Phys. Rev. Lett.* **114,** 257202 (2015).

25. Qiu, Z. *et al.* Spin-current probe for phase transition in an insulator. *Nat. Commun.* **7,** 12670 (2016).

26. Shigematsu, E. *et al.* Significant reduction in spin pumping efficiency in a platinum/yttrium iron garnet bilayer at low temperature. *Appl. Phys. Express* **9,** 53002 (2016).